\documentclass[conference, a4paper,cmex10]{IEEEtran}
\IEEEoverridecommandlockouts

	\usepackage[turkish]{babel}
	\usepackage[utf8]{inputenc} 
	\usepackage[T1]{fontenc}

\usepackage{xcolor}
\usepackage{amsmath,graphicx}
\usepackage{cite}
\usepackage{amssymb,amsfonts}
\usepackage{algorithmic}
\usepackage{graphicx}
\usepackage{textcomp}
\usepackage{xcolor}
\usepackage{booktabs}
\usepackage{cite}
\usepackage{amsmath}
\usepackage{multirow}
\usepackage{array}
\usepackage[lofdepth,lotdepth]{subfig}

\AtBeginDocument{%
  \renewcommand\tablename{TABLO}
}

\AtBeginDocument{%
  \renewcommand\abstractname{Abstract}
}

\begin{document}

\IEEEpubid{\makebox[\columnwidth]{978-1-6654-3649-6/21/\$31.00 ©2021 IEEE\hfill}
\hspace{\columnsep}\makebox[\columnwidth]{}}

%
\title{Sınırlı Kaynaklı Ortamlar için Yapay Sinir Ağları Bazlı Uyku Fazları Sınıflandırılması \\
Neural Network Based Sleep Phases Classification for Resource Constraint Environments}


%

\author{\IEEEauthorblockN{ Berkay Köprü}
\IEEEauthorblockA{Koç Üniversitesi \\
İstanbul, Türkiye \\
bkopru17@ku.edu.tr}
\and
\IEEEauthorblockN{Murat Aslan, Alisher Kholmatov}
    \IEEEauthorblockA{Maxim Integrated \\
    İstanbul, Türkiye \\
    \{murat.aslan, alisher.kholmatov\}@maximintegrated.com}
}


%

\maketitle

\begin{ozet}
Uyku vücudun kendini yenilediği bir süreç olup, bu yenilenmenin efektifliği ise uyuyanın belli uyku fazlarında ne kadar süre geçirdiğiyle doğrudan ilintilidir. Bu sebeple, giyilebilir teknolojiler ile uyku değerlerinin takip edilmesi hem araştırmacılar hem de endüstri tarafından ilgi çekici bir konudur. 
Mevcut uyku takibi çözümlerinin en gelişmiş olanlarının işlemci hafızası ve işlem kullanımı yüksek olduğu gibi aynı zamanda da bulut veya cep telefonu bağlantısına ihtiyaçları vardır. Bu çalışmada tamamen gömülü ortamda çalışabilien ve bulut veya mobil telefon bağlantısına ihtiyaç duymayan hafıza açısından efektif bir mimari önerilmiştir. 
Çalışma kapsamında, uyku fazlarının sınıflandırılması için öznitelik çıkarımı ve Yapay Sinir Ağları bazlı bir istifleme sınıflandırmasından oluşan özgün bir mimari önerilmiştir. Önerilen yöntemi doğrulamak amacıyla 24 farklı kişiden 31 gecelik veri kümesi oluşturulmuştur. Bu veri kümesi 3 eksen ivme-ölçer (ACC) ve fotopletismogram (PPG) sensörleriyle donatılmış bileğe takılan bir cihaz ile toplanmıştır. Bu veri seti üzerinde çapraz doğrulama ile yapılan deneyler sonucu uyku fazı sınıflandırmasında ise karşılaştırılan sistemlerden F1 açısından \%20 ve \%14 daha iyi sonuç elde edilmiştir. Sistemin üstün performansının yanında, hafızada (RAM) sadece 4.2 kilobaytlık yer kaplıyor olması, sınırlı kaynaklı gömülü sistemler için çok uygun olduğunun göstergesidir.

\end{ozet}
\begin{IEEEanahtar}
Nabız, Kalp Hızı Değişkenliği, Uyku Fazı Sınıflandırması, Yapay Sinir Ağları, Giyilebilir Cihazlar.
\end{IEEEanahtar}

\begin{abstract}
Sleep is restoration process of the body. The efficiency of this restoration process is directly correlated to the amount of time spent at each sleep phase. Hence, automatic tracking of sleep via wearable devices has attracted both the researchers and industry. Current state-of-the-art sleep tracking solutions are memory and processing greedy and they require cloud or mobile phone connectivity. We propose a memory efficient sleep tracking architecture which can work in the embedded environment without needing any cloud or mobile phone connection.
In this study, a novel architecture is proposed that consists of a feature extraction and Artificial Neural Networks based stacking classifier. Besides, we discussed how to tackle with sequential nature of the sleep staging for the memory constraint environments through the proposed framework. To verify the system, a dataset is collected from 24 different subjects for 31 nights with a wrist worn device having 3-axis accelerometer (ACC) and photoplethysmogram (PPG) sensors. Over the collected dataset, the proposed classification architecture achieves 20\% and 14\% better F1 scores than its competitors. Apart from the superior performance, proposed architecture is a promising solution for resource constraint embedded systems by allocating only 4.2 kilobytes of memory (RAM).
%
\end{abstract}
\begin{IEEEkeywords}
Heart Rate, Heart Rate Variability, Sleep Phase Classification, Artificial Neural Networks, Wearable Devices.
\end{IEEEkeywords}



%
\IEEEpeerreviewmaketitle

\IEEEpubidadjcol

\section{G{\footnotesize İ}r{\footnotesize İ}ş}
Uyku, insanların beden ve zihinsel sağlığı için çok önemlidir \cite{long2013sleep}.  Uyku sırasında, beden aktif olarak hafızanın derlenmesi ve yeniden yapılandırılması üzerinde çalışır. Uykusuzluğun, zihinsel yeteneklerdeki kayıpların yanında, kalp damar hastalıklarına da yol açtığı tespit edilmiştir \cite{he2017association}.

Hipokampüs, uyku döngüsünü sirkadiyen ritmin bir parçası olarak kontrol etmektedir. Kontrol edilen uyku döngüsü, hızlı göz hareketleri (REM) ve hızlı olmayan göz hareketleri (NREM) olarak iki temel evre arasında döngüsel olarak değişmektedir. Ayrıca, NREM fazı kendi içinde derin (DEEP/NREM 3-4) ve hafif (LIGHT/NREM 1-2) olarak ikiye ayrılır. Ayrıca LIGHT fazı DEEP ile REM arasındaki bir geçiş fazı olarak adlandırılır.

Uyku fazlarını tespit etmek için altın standart  polisomnografi (PSG) yöntemidir. PSG yönteminin uygulanması sırasında, hastadan  electroensefalografi (EEG), elektrookulografi (EOG), elektromyelografi (EMG), elektrokardiyografi (EKG), kapnografi ve hareket sinyalleri toplanır. Toplanan sinyaller 30 saniyelik ayrık parçalara bölünerek yetkili birimler tarafından yorumlanır ve bu parçalar epok olarak adlandırılır. PSG yönteminin pahalı olması, çok geniş laboratuvar olanakları gerektirmesi ve birçok cihazın aynı anda hastaya bağlanması sebebiyle hastayı uyku sırasında rahatsız etmesi gibi sorunları vardır. PSG'nin oluşturduğu sorunları çözebilmek adına uyku takibinin kardiyorespiratuvar sinyalleri takip edebilen giyilebilir cihazlarla yapılması önerilmiştir \cite{sleep_stage_classification_multi_level_2017}. 

Bu araştırmanın ana katkıları aşağıdaki gibidir:

\begin{itemize}
    \item Doğal uykunun takibi için, PPG ve ivme ölçer verileri içeren özgün bir veri kümesi toplanması
    \item Uyku fazlarının döngüsel yapısını açığa çıkarabilen nabız ve kalp atım aralığı bazlı öznitelik çıkarımı önerimi
    \item Sınırlı kaynaklı gömülü ortamlarda çalışabilecek karmaşıklıkta olan istifleme ve yapay sinir ağları bazlı sınıflandırma yöntemi önerimi
\end{itemize}

\subsection{İlg{\footnotesize İ}l{\footnotesize İ} Çalışmalar}

 EEG bazlı uyku fazı sınıflandırılması çalışmaları \cite{phan2019seqsleepnet_eeg}, \cite{mikkelsen2018personalizing_eeg} gömülü ortamların kısıtlarını dikkate almamışlardır. Mikkelsen vd. \cite{mikkelsen2018personalizing_eeg} EEG, EOG ve EMG sinyallerini zaman-frekans imgelerine çevirip önce evrişimli sinir ağları (CNN) benzeri katmana daha sonra da iki katmanlı tekrarlayan sinir ağlarına (RNN) aktarmıştır. \cite{mikkelsen2018personalizing_eeg} ise kişisel bilgileri kullanarak 17 milyon parametreli yaklaşık 68 megabaytlık bir çözüm sunarak Uyanıklık (WAKE), REM, NREM1, NREM2 ve NREM3 olmak üzere 5 sınıf sınıflandırmada 0.87 doğruluk performansına ulaşmıştır.

Uyku araştırmalarındaki bir diğer eğilim ise, yüksek örnekleme hızına sahip EKG sensörünü kullanarak kalp hızı değişkenliği (HRV) bazlı özniteliklerle uyku analizi yapmaktır \cite{mendez2010sleep_ecg}. Mendez vd. \cite{mendez2010sleep_ecg} çalışmalarında 128 Hz ile toplanmış kalp atım aralığı sinyalinden, zaman ve frekans bazlı HRV öznitelikleri çıkarıp Saklı Markov Modelleri ile REM, NREM, ve Uyanıklık (WAKE) olmak üzere 3 sınıflı sınıflandırmasını 0.79 doğruluk ile gerçeklemişlerdir. Ancak yüksek örnekleme hızı ve frekans düzleminde öznitelik çıkarmak gömülü ortamlar için hafıza yetersizliği sıkıntısına yol açmaktadır.

\cite{zhang2018sleep_ppg} ve \cite{fonseca_main_comparison_2017}, PPG ve ivme ölçer sensörlerini kullanarak uyku fazı sınıflandırmasını gerçeklemişlerdir. Zhang vd. \cite{zhang2018sleep_ppg} giyilebilir saatle topladıkları nabız bilgisinden HRV öznitelikleri çıkarıp çok düğümlü Uzun-Kısa Vadeli Bellek (LSTM) ile 5 sınıflı sınıflandırmasında 0.6 doğruluğa ulaşmışlardır. Ancak bu çalışma öznitelik çıkarımında kullanılan Ayrık Kosinüs Dönüşümü (DCT) ve sınıflandırma için kullanılan çok katmanlı ve karmaşıklığı yüksek olan LSTM kullanılması açısından sınırlı kaynaklı ortamlar için uygun değildir.  \cite{fonseca_main_comparison_2017} uyku fazlarini siniflandirabilmek icin kalp atım aralıklarından Kalp Hızı Değişkenliği (KHD) öznitekleri çıkarmıştır. Zaman ve freakans alanında hesaplanan KHD öznitelikleri, ivme ölçerden hesaplanan öznitekliklerle birleştirilerek öznitelik seviseyinde bir füzyon önerilmiştir. Uyku fazlarının sınıflandırılması çok sınıflı Bayesçi doğrusal ayırtaç ile yapılmıstır.

\section{Yöntem}

Bu bölümde uyku takibi için çıkarılan öznitelikler ve sınıflandırma tanımlanacaktır. Önerilen sistemin ilk adımı nabız ve kalp atım aralıkları sinyallerinden istatistiksel ve aritmetiksel öznitelikler çıkarmaktır. Çıkarılan özniteliklerle Yapay Sinir Ağları (ANN) eğitip, uyku fazları sınıflandırılması için kullanılmıştır.

\begin{figure}[h]
	\centering
	\shorthandoff{=}  
	\includegraphics[width=\columnwidth]{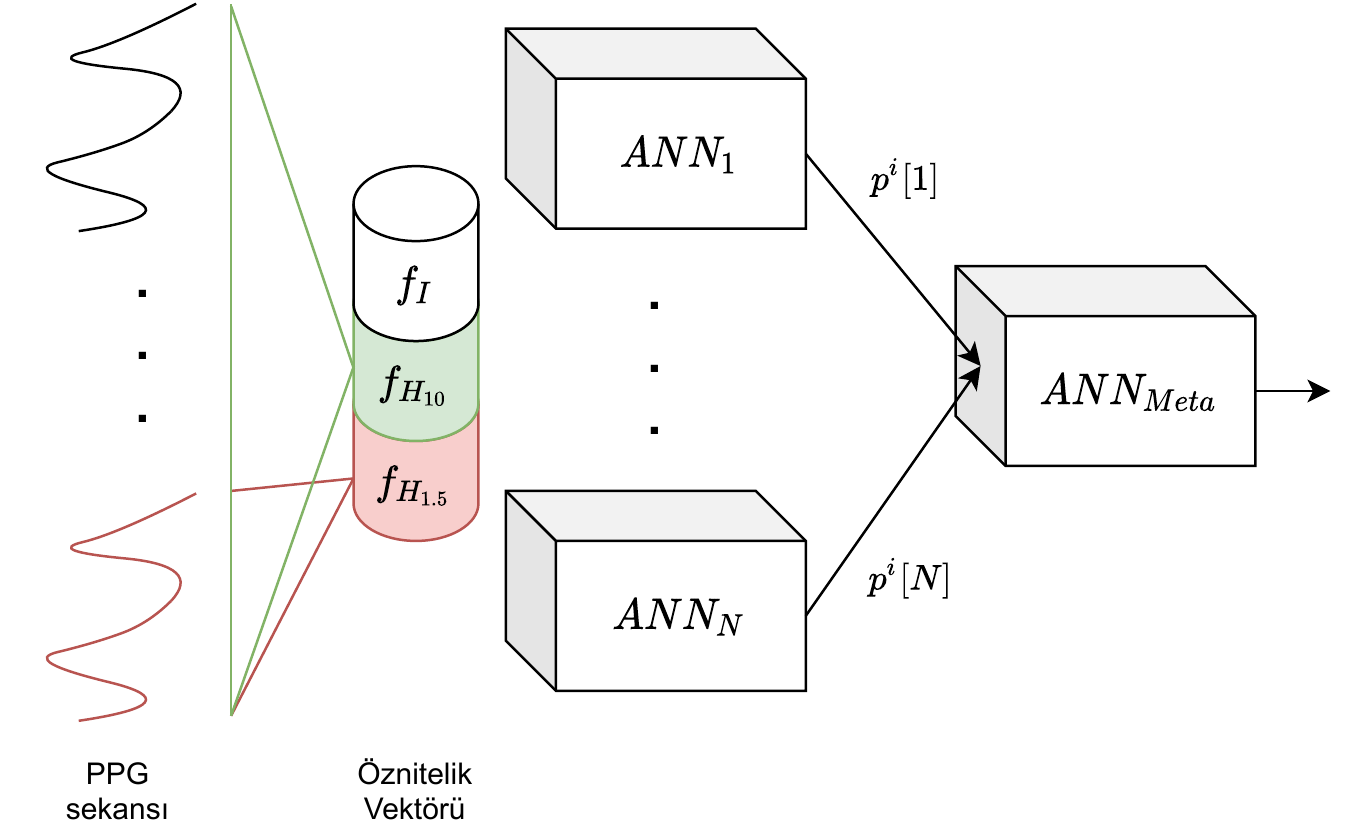}
	\shorthandon{=} 
	\caption{ANN kullanarak önerilen istifleme mimarisi}
	\label{fig::stacking}
\end{figure}


\subsection{Öznitelik Çıkarımı}

Önerilen sistemin girdileri olarak nabız ve kalp atım aralığı kullanılmaktadır ve bu girdiler EKG yerine PPG'den hesaplanmış sinyallerdir. Anlık nabız ve kalp atım aralığının uyku fazlarına etkisinin incelenmesi için belli bir uzunluktaki pencere üzerinden bakılmalıdır. Çalışmamızda iki farklı pencereden çıkarılmış öznitelikler birleştirilip sınıflandırma yöntemine verilmiştir. Özellikle sistemin karmaşıklığını arttırmamak için zaman alanından öznitelik çıkarılmıştır. 

\subsubsection{Nabızdan Öznitelikleri}

Bu çalışmada nabız örnekleme oranı ($f_s$) 25 Hz olarak kullanılmıştır, ve Şekil 1'de görüldüğü gibi 1.5 ve 10 dakikalık iki pencere üzerinden öznitelik çıkarılmıştır. Depolanan nabız vektöründen öznitelik olarak standart sapma, ortalama, minimum, maksimum ve 25. ile 75. persentilleri çıkarılıp $f_{H_{t}}$ vektörü seklinde ifade edilmiştir ve $t \in \{1.5,10\}$.


\subsubsection{Kalp Hızı Değişkenliği Öznitelikleri}

Nabızdan ayrı olarak kalp atım aralıkları düzensiz gelmektedir, ve kalp atım aralıklarından çıkarılan özellikler KHD olarak adlandırılır. Depolanan kalp atım aralıklarından değişim katsayısı, standart sapma, ortalama öznitelik olarak kullanılmıştır. Kalp atım aralıklarının farklarından ise, standart sapma, karelerinin ortalamasının kökü (RMSSD), 50 milisaniyeden fazla olan farkların sayısı (NN50) ve toplam vektörün uzunluğuna oranı (PNN50), 20 milisaniyeden fazla olan farkların sayısı (NN20) ve toplam vektörün uzunluğuna oranı (PNN20) bazlı olmak üzere $f_{I}$ vektörü oluşturulmuştur. 
%
%
\begin{table}[ht]
\begin{center}
\begin{tabular}{l c  c  c c}
\toprule[1pt]\midrule[0.3pt]
    \textbf{Nabız (1.5 dk)}       & \textbf{Nabız (10 dk)} & \textbf{Kalp Hızı Değişkenliği} \\  \midrule \hline
    Ortalama (1) &       Ortalama (1) &           Değişim katsayısı (1)\\ 
    Standart Sapma (1) & Standart Sapma (1) &     Standart Sapma (2)\\ 
    Minimum (1) &        Minimum (1) &            RMSSD (1)\\ 
    Maksimum (1) &       Maksimum (1) &           NN50 (1)\\ 
    Persentil (2)&       Persentil (2)&           PNN50 (1) \\ 
     &                                &           NN20 (1) \\ 
     &                                &           PNN20 (1) \\ 
     &                                &           Ortalama (2) \\

\end{tabular}         
\end{center}
 \caption{Önerilen mimarinin girdisi olarak nabız ve kalp atımından çıkarılan öznitelikler}
    \label{tab:features}
\end{table}
Sonuç olarak sınıflandırma için kullanılacak öznitelik vektörü $\mathbf{f} \in \mathbb{R}^{22\times1}$,
\begin{equation}
    \mathbf{f} = \{f_{I},f_{H_{10}},f_{H_{1.5}}\},
\end{equation}
$f_{I}$, $f_{H_{10}}$ ve $f_{H_{1.5}}$'nin yan yana eklenmesiyle oluşturulmuştur. Özniteliklerin detayları Tablo~\ref{tab:features}'de ayrıntılı şekilde gösterilmiştir.

\subsection{Sınıflandırma}

Sınıflandırma sistemi her bir $\mathbf{f}$ için 3 sınıflı $R=\{REM, Light, Deep\}$ kümesinden birini seçerek bir karar verir. Uyku fazları ile nabız ve kalp hızı değişkenliğinin lineer olmayan ilişkisi göz önüne alinarak uyku fazları sınıflandırılması amacıyla ANN kullanılmıştır. 

\subsection{İstifleme}

İstifleme sistemi düşük katman sayısına sahip ANN’leri paralel olarak ekleyerek, eğitim sisteminin parametre sayısını lineer olarak arttırmaya dayalı bir çözümdür. 

\begin{figure}[h]
	\centering
	\shorthandoff{=}  
	\includegraphics[scale=0.2]{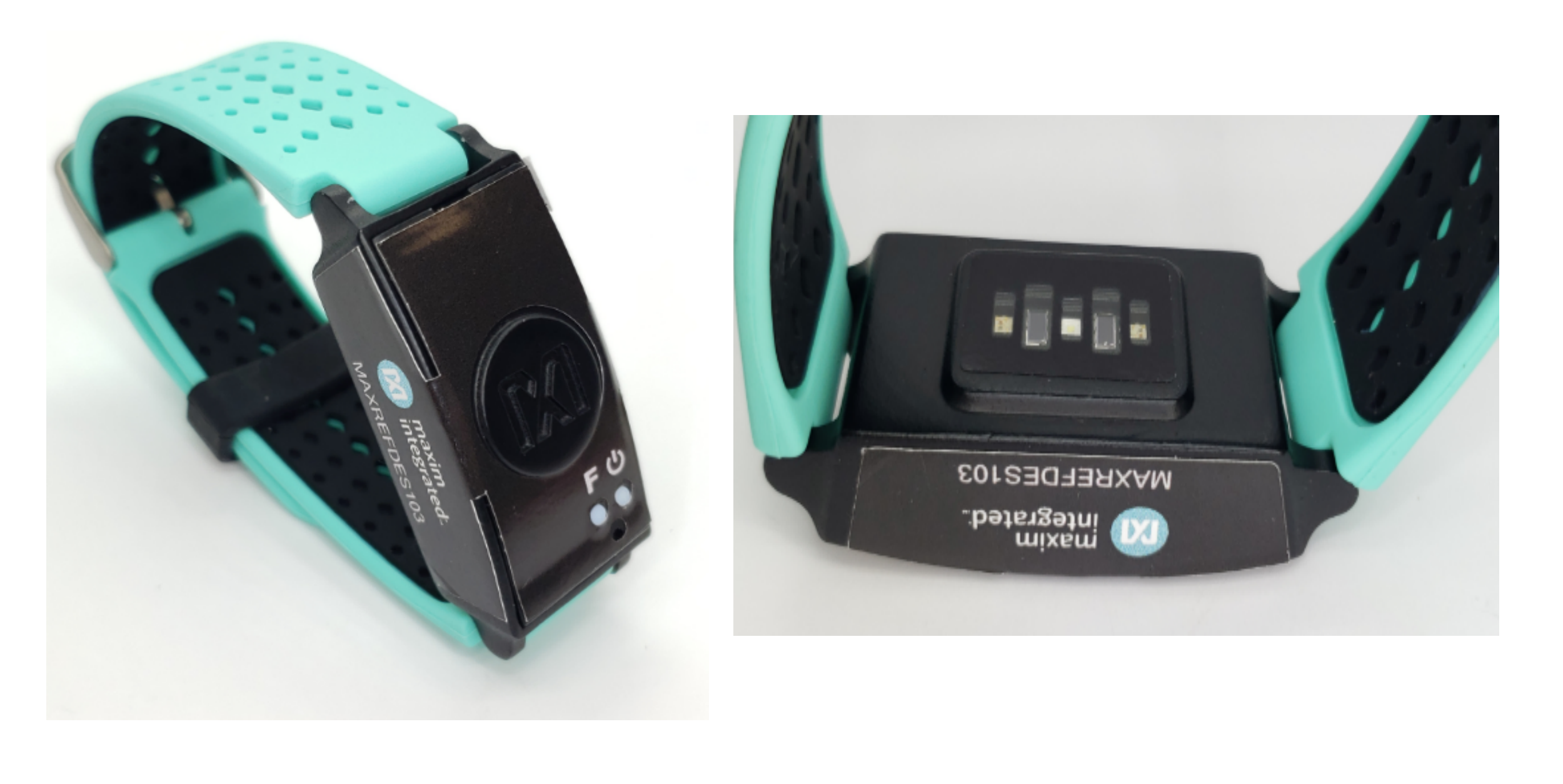}
	\shorthandon{=} 
	\caption{MRD103'ün görüntüleri}
	\label{fig::mrd103}
\end{figure}

Bu çalışmada önerdiğimiz istifleme bazlı mimari Şekil \ref{fig::stacking}'de gösterilmiştir. Bu mimari iki ana modulden oluşmaktadır: $N$ tane parelel baz ANN, ve meta ANN. Baz kısımdakı her bir ANN, uyku fazları tahmini için $\mathbf{f}$ ile egitilirken, ikinci modül meta-ANN baz ANN'lerin tahminleriyle ($p^{i}$) uyku fazları tahmini için eğitilir. Böylece meta-ANN, baz ANN'lerin hangisine güveneceğini, ve tahminlerin en iyi şekilde nasıl birleştirilip uyku fazı tahmini yapılacağını öğrenmiş olur.


\section{Deneysel Çalışma}
\begin{figure*}[h]
	\centering
	\shorthandoff{=}  
	\includegraphics[width=10.5cm,height=3cm]{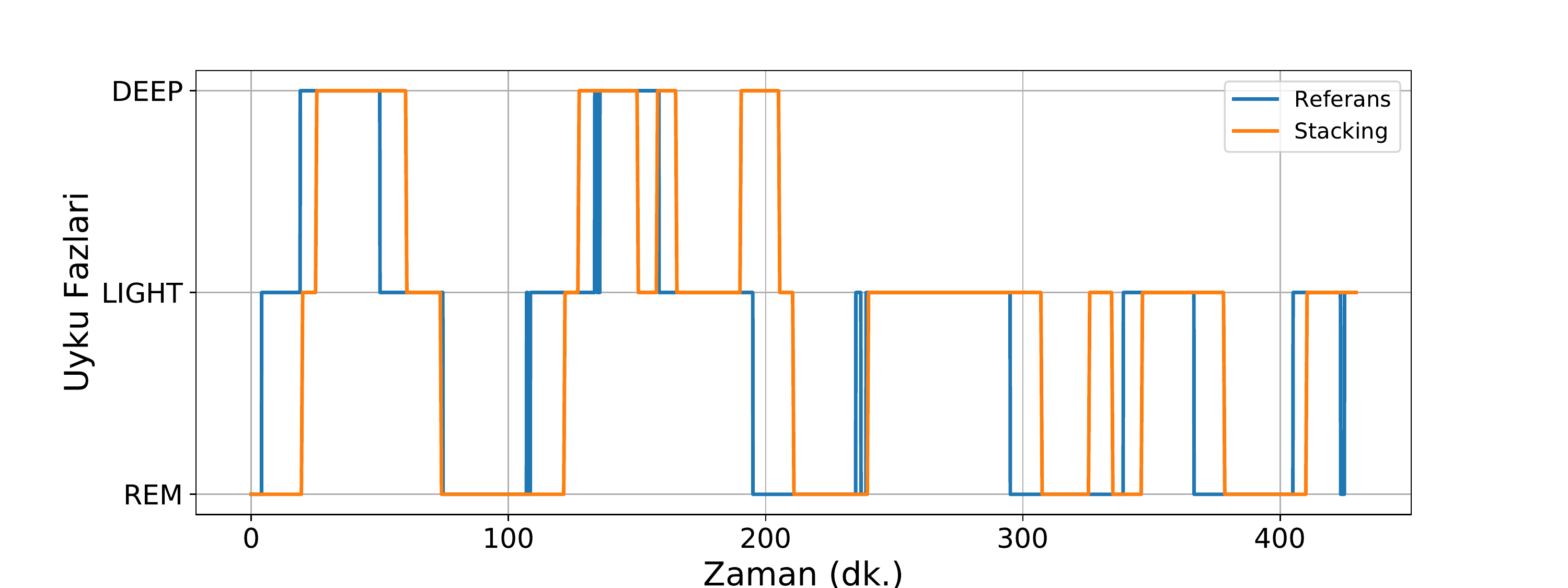}
	\shorthandon{=} 
	\caption{İstifleme yönteminin bir gecelik veri üzerinde referans ile karşılaştırılması  }
	\label{fig::stacking_vs_ref}
\end{figure*}
Şekil \ref{fig::stacking}'de önerilen mimarinin eğitimi sırasında 0.001 öğrenme hızıyla Adam en iyileyici kullanılmıştır. Öğrenme sırasında 32'lik yığınlar (batch) kullanılmış olup, 10 epokluk sabırlı erken durdurma (Early Stopping) seçeneğiyle 50 epok boyunca eğitim yapılmıştır.

Önerilen mimarideki ANN'ler 2 katmanlı olup, katmanlar arasında 0.2'lik bırakma (Drop-Out) katmanı kullanılmıştır. Birinci katmanda 9 nöron, ikinci katmanda 7 nöron kullanılmıştır. Çalışmamızda $\text{N}=4$ olarak kullanıp, iki ANN'i \textit{RELU} aktivasyonuyla, diger iki ANN ise \textit{TanH} aktivasyonuyla eğittik. Meta ANN istiflenen sinir ağlarından farklı olarak tek katmanlı olup, 5 nöron ve lineer aktivasyona sahiptir. 

\subsection{Veri Kümesi}

Çalışma kapsamında 24 farklı kişiden 31 gecelik uyku verisi toplanmıştır. Bu veriler Şekil~\ref{fig::mrd103}'de gösterilen, çalışma grubu tarafından referans tasarım olarak geliştirilmiş, bir giyilebilir saat (MRD103) ile toplanmıştır. MRD103 içinde fotopletismografi (PPG) ve ivme ölçer bulunduran bir saat olmakla beraber, bu sensörler saati takan insanın hareketlerini ve nabzını takibini sağlamak için kullanılır.

Veri toplama sırasında deneklerden MRD103'u bileklerine takmaları ve saati uykuya dalmadan en az 30 dakika önce takmış olmaları istenmiştir. MRD103 ile beraber denekler referans cihazı olarak kullandığımız EEG bazlı kafa bandını (Zeo Sleep Manager) takmaları istenmiştir. Toplanan veri kümesinin istatistik bilgileri Tablo~\ref{tab:corr_mse} verilmiştir.

\subsection{Değerlendirme Kriteri}

Bu çalışma temelde bir sınıflandırma olduğu için, performans değerlendirmesini ağırlıksız ortalama Doğruluk(D) ve F1-skoru kriterleri üzerinden yapılmıştır. Ortalama performans değerlendirmesi, 
her bir sınıfın performansını kullanarak aşağıdaki gibi hesaplanmıştır:

\begin{align}
    UD  &= \frac{1}{N} \sum_{i}^{N}D_{i}, \\
    UF1 &= \frac{1}{N}\sum_{i}^{N} F1_{i}
\end{align}

ve $i \in \{REM,Light, Deep\}$, $N$ sinif sayisi ($N=3$) olarak adlandırılmıştır.

\begin{table}[htb]
\begin{center}
\begin{tabular}{l c c c}
\toprule[1pt]\midrule[0.3pt]
   & \textbf{Ortalama}   & \textbf{Standart Sapma}    \\  \midrule \hline
\textbf{Yaş} & 31.05       & 4.53   \\ \hline
\textbf{Deney Süresi}      & 388.10 dk       & 114.13 dk       \\ \hline
\textbf{Uyku Süresi}      & 303.57 dk       & 116.24 dk    \\ \hline
\textbf{Uyaniklik Süresi}      & 84.51 dk       & 59.77 dk    \\ \hline
\textbf{REM Süresi}      & 81.97 dk       & 41.23 dk    \\ \hline
\textbf{Light Süresi}      & 158.51 dk       & 72.06 dk    \\ \hline
\textbf{Deep Süresi}      & 60.41 dk      & 35.68 dk   \\ \hline
\end{tabular}
\end{center}
 \caption{Uyku takibi çalışması kapsamında toplanan veri kümesinin istatistik bilgileri}
    \label{tab:corr_mse}
\end{table}

\subsection{Sonuçlar}

\begin{figure}[h]
	\centering
	\shorthandoff{=}  
	\includegraphics[scale=0.45]{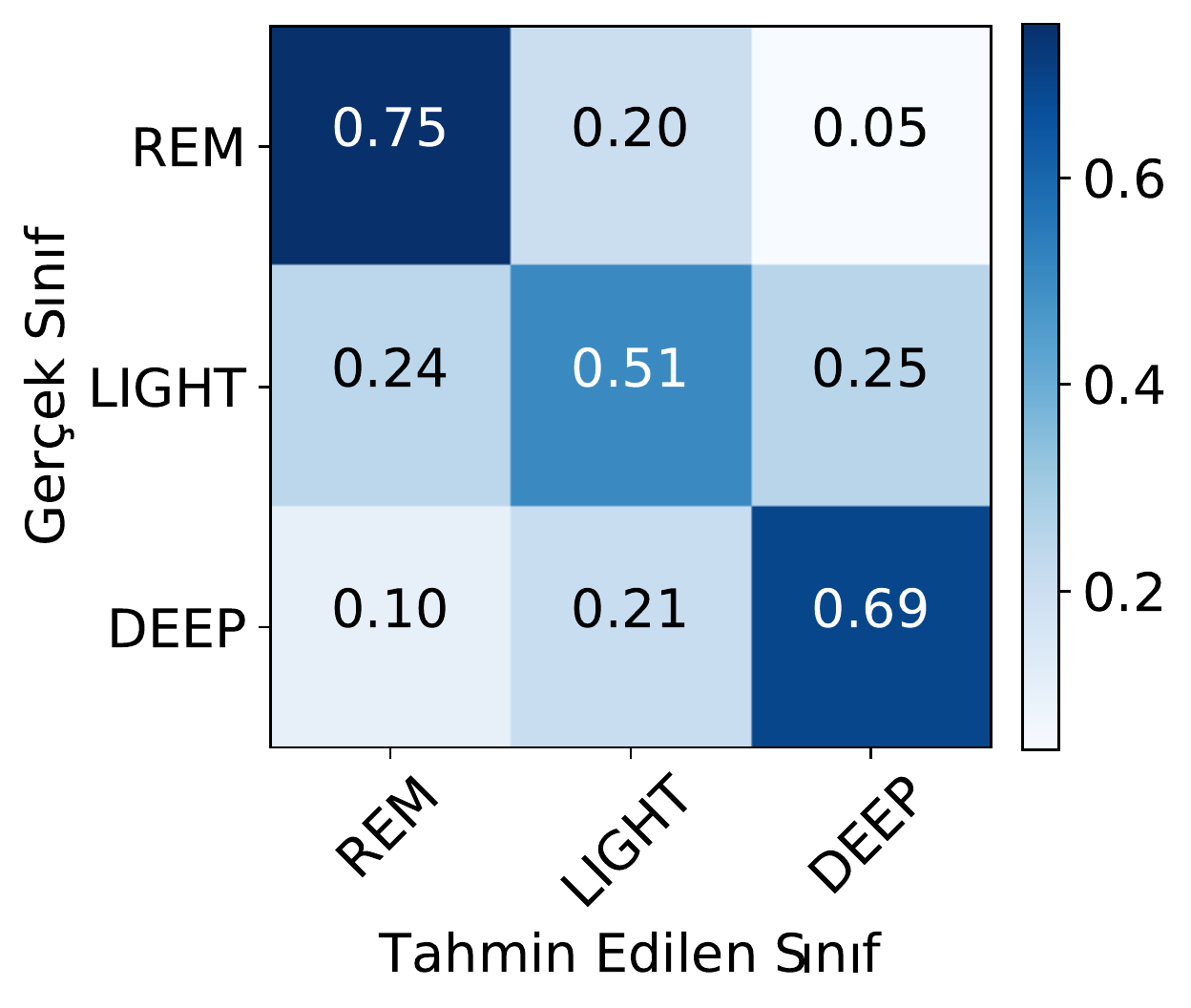}
	\shorthandon{=} 
	\caption{Önerilen istifleme metodunun performansı}
	\label{fig::conf_mtx}
\end{figure}

Şekil~\ref{fig::conf_mtx}'de görüldüğü üzere istifleme metodu REM ve DEEP sınıflarını sınıflandırmada üstün bir performansa sahiptir. REM ve DEEP süreleri uykunun kalitesine LIGHT sınıfından daha çok etki eder, ve bu sınıfların doğru şekilde belirlenişi önemlidir. Sistemin REM'deki 0.75'lik duyarlılık başarısı, ise insan vücudunun REM fazında uyanıklığa benzer hareket etmesi, dolayısıyla nabızdan çıkarılan özniteliklerin diğer uyku fazlarından ayrık olmasıyla açıklanabilir. Önerilen metot LIGHT sınıflandırması için 0.51 duyarlılığına sahip ve diğer sınıflara bir yanlılığı olmadığı görülmüştür. Bu durum LIGHT evresinin REM ile DEEP arasındaki bir geçiş evresi olmasıyla açıklanabilir.

\begin{table}[ht]
\begin{center}
\begin{tabular}{l c  c  c c}
\toprule[1pt]\midrule[0.3pt]
    \textbf{Model}       & \textbf{Doğruluk(UD)} & \textbf{F1 Skoru(UF1)} &\textbf{Parametre Sayısı}\\  \midrule \hline
    ANN${0}$ & 0.51 & 0.49 & 411\\ \hline 
    ANN${1}$ & 0.56 & 0.48 & 411\\ \hline 
    ANN${2}$ & 0.49 & 0.48 & 411\\ \hline 
    ANN${3}$ & 0.53 & 0.53 & 411\\ \hline 
    ANN BIG & 0.54 & 0.5 & 2200\\ \hline
    LSTM & 0.56 & 0.53 & 8623\\ \hline
    \textbf{STACKING}  & \textbf{0.60} & \textbf{0.61} & 1720\\ \hline
\end{tabular}
\end{center}
 \caption{Farklı sınıflandırma yöntemleriyle önerilen istifleme yaklaşımının 3 sınıflı sınıflandırması için performans karşılaştırılması}
    \label{tab:performance}
\end{table}

İstifleme yöntemi, Şekil 1'deki baz ANN'ler, 6 katmanlı (ANN BIG) bir yapay sinir ağı ve LSTM ile karşılaştırılmıştır. İstifleme yöntemi
Tablo~\ref{tab:performance}'de gösterildiği gibi hem Doğruluk hem de F1 Skorunda diğer mimarilerden daha üstün bir performansa sahiptir. İstifleme metodu baz olarak kullanılan ANN'lerden Doğruluk'ta en az \%8, F1-skorunda en az \%14 daha başarılıdır. Bu sonuçlar meta-ANN ile ikinci bir eğitim basamağının başarısını göstermektedir. Özellikle önerilen sistemin, 2 katman ve 25 üniteli LSTM'den F1 açısından \%14 daha iyi sonuç elde etmesi dikkat çekicidir. LSTM'lerin ANN'lerden yüksek karmaşıklıkları dikkate alınğında, daha üstün bir performans sağlamaları beklenir. Önerilen sistemin bir gecelik veri üzerinde çıktıları ve referans ile karşılaştırması Şekil~\ref{fig::stacking_vs_ref}'te örnek olarak verilmiştir.


Sistemin performansını literatürle karşılaştırabilmek adına sistemi Uyanıklık (WAKE), REM ve NREM (LIGHT-DEEP) şeklinde 3 sınıflı sınıflandırma için tekrar eğittik. Ancak biz bu çalışmada öznitelikleri çıkarırken,  \cite{fonseca_main_comparison_2017}'nin aksine denek bazlı bir öznitelik ölçeklendirmesi yapmayıp, bütün deneklerin öznitelikleri üst üste eklendikten sonra öznitelik ölçeklendirilmesi yaptık. Denek bazlı ölçeklendirme izlendiğinde önerdiğimiz mimarinin doğruluğunun 0.75'e ulaştığı gözlemlenmiştir, ve \cite{fonseca_main_comparison_2017}'nin önerdiği ve doğruluğu 0.73 olan sistemden daha üstün olduğu görülmüştür. Bu sonuç önerilen mimarinin aynı koşullar altında literatürden daha iyi bir başarımının olduğunu göstermektedir.

Önerdiğimiz sınıflandırma yönteminin parametre sayısı 1720 olup, herhangi bir sıkıştırma yöntemi olmaksızın 4 baytlık (byte) reel sayılarla (float) gerçeklendiğinde, gömülü ortamda yaklaşık 4.2 kilobaytlık (kB) bir alan kaplar. Örnek bir mikrodenetleyici olan MAX32660 96 kB'lık RAM'e sahiptir. Önerilen yöntem MAX32660 gibi tipik bir mikrodenetleyicinin RAM'inin sadece $\%4.4$'ünü işgal ederek, sınırlı kaynaklı ortamlar için hem yüksek performanslı hem de hafıza açısında hesaplı bir çözüm olduğunu göstermiştir.

Önerilen sistemin aynı koşullarda literatürden daha iyi doğruluğa sahip olması, ve bu performansa düşük alan ve enerji ihtiyaçlarıyla ulaşması, bu çalışmanın uzun süreli uyku takibini mümkün kıldığı göstermiştir.

\section{Sonuç}

Bu çalışmada nabız ve kalp atım aralıkları bazlı öznitelik çıkarımı ile istifleme bazlı bir sınıflandırma sistemi önerilmiştir. Bu özgün öznitelik çıkarım yöntemi uyku fazlarının döngüsel yapısını göz önüne alarak farklı zaman pencereleri üzerinde işlem yapmaktadır. Çıkarılan öznitelikler istifleme yöntemiyle birleştirilmiş ANN'lere aktarılarak sınıflandırılma gerçekleştirilmiştir. 

Önerilen sistemin efektifliğini gösterebilmek amacıyla 24 farklı kişiden 31 gecelik bir veri kümesi toplanmıştır. Toplanan veri kümesi üzerinde çapraz doğrulama yöntemiyle önerilen sistemin F1 ve doğruluk performansı hesaplanmıştır. Bu deneyler sonucu görülmüştür ki F1 açısından ANN BIG'ten \%20, RNN'den \%14  daha iyi sonuçlar elde etmiştir. Ayrıca, bu performansı ortalama bir mikro denetleyicinin hafızasının sadece \%4’ünü kapsayarak elde etmiştir




%

\bibliography{bare_conf.bib}
\bibliographystyle{IEEEtran}

\end{document}